\documentclass[superscriptaddress,showpacs,twocolumn,amssymb,aps]{revtex4-1}

\usepackage{graphicx,dcolumn,bm,amsmath,color}

\newcommand{\eq}[1]{Eq.~(\ref{#1})}
\newcommand{\heq}[1]{Eq.~(\ref{#1})}
\newcommand{\fig}[1]{fig.~\ref{#1}}

\newcommand{\eeq}{ \end{equation} }
\newcommand{\beq}{ \begin{equation} }

\newcommand{\eea}{ \end{eqnarray} }
\newcommand{\bea}{ \begin{eqnarray} }

\newcommand{\oma}{\Omega_{1}}

\newcommand{\bs}{ {\bf s }}

\newcommand{\bhua}{ {\bf \hat{u}}_{1} }
\newcommand{\bhub}{ {\bf \hat{u}}_{2} }

\newcommand{\bhu}{ {\bf \hat{u}} }
\newcommand{\az}{  \psi}
\newcommand{\aza}{  \psi_{1} }
\newcommand{\azb}{  \psi_{2} }
\newcommand{\ta}{  t_{1} }
\newcommand{\tb}{  t_{2} }

\newcommand{\bfr}{ {\bf r} }

\newcommand{\bn}{ {\bf \hat{n}} }

\newcommand{\bv}{ {\bf \hat{v}} }
\newcommand{\bw}{ {\bf \hat{w}} }

\begin{document}

\title{Spontaneous sense inversion in helical mesophases}

\author{H. H. Wensink}
\email{wensink@lps.u-psud.fr}

\affiliation{Laboratoire de Physique des Solides, Universit\'e Paris-Sud  \& CNRS, UMR 8502, 91405 Orsay, France}

\date{\today}

\begin{abstract}
We investigate the pitch sensitivity of cholesteric phases of helicoidal patchy cylinders
as a generic model for chiral (bio-)polymers and helix-shaped colloidal rods.
The behaviour of the macroscopic cholesteric pitch  
is studied from microscopic principles by invoking a simple density functional theory generalised to accommodate weakly twisted
director fields. 
Upon changing the degree of alignment along the local helicoidal director we find that cholesteric phases exhibit a sudden sense inversion whereby the cholesteric phase changes from left- to
right-handed and vice versa. Since the local alignment is governed by thermodynamic variables such as density,  temperature or the amplitude of an external directional field such pitch sense inversions can be expected in systems of helical mesogens of both thermotropic and lyotropic origin. We show that the spontaneous change of helical symmetry  is a direct consequence of an antagonistic effective torque between helical particles with a certain prescribed internal helicity.  The results may help opening up new routes towards  precise control of the helical handedness of chiral assemblies by a judicious choice of external control parameters.
\end{abstract}


\maketitle

Over the past decades considerable research effort has been devoted to
understanding the manifestation of macroscopic chirality in lyotropic liquid crystals consisting of
colloidal particles or stiff polymers immersed in a solvent. 
In addition to a number of synthetic helical polymers such as polyisocyanates \cite{aharoni,sato-sato} and polysilanes \cite{watanabe-kamee} which form cholesteric phases in organic solvents there is a large class of helical bio-polymers which are known to form cholesteric phases in water. Examples are  DNA
\cite{robinson-pblg,livolantDNAoverview} and the rod-like $fd$-virus
\cite{dogic-fraden_fil}, polypeptides \cite{uematsu,dupresamulski}, chiral micelles \cite{hiltrop},
polysaccharides \cite{sato-teramoto}, and microfibrillar cellulose derivatives
\cite{gray-cellulose} and chitin \cite{chitin-revol}. In these systems, the
cholesteric pitch is strongly dependent upon the particle concentration, temperature as well as solvent properties such as the ionic strength. The effect of these individual factors on the macroscopic pitch  has been the subject of intense experimental research \cite{robinson-pblg,rill-dna,yu-dna,strey-dna, dogic-fraden_chol,grelet-fraden_chol,dupreduke75,yoshiba-sato,dong-gray-cellulose,miller-cellulose,microcellulose}.

The connection between the molecular interactions responsible for
chirality on the microscopic scale and the structure of the
macroscopic cholesteric phase is very subtle and has been a long-standing challenge in the physics
of liquid crystals \cite{gennes-prost}.  The chiral nature of most biomacromolecules originates from a  spatially
non-uniform distribution of charges and dipole moments residing on the
molecule. The most prominent example is the double-helix backbone structure of the phosphate
groups in DNA.  Combining the electrostatic interactions with the intrinsic
conformation of the molecule  allows for a coarse-grained
description in terms of an {\em effective} chiral shape. 
Examples are bent-core or banana-shaped
molecules \cite{straley,jakli} where the mesogen shape is primarily responsible for
chirality. Many other helical bio-polymers and microfibrillar
assemblies of chiral molecules (such as cellulose) can be mapped onto effective
chiral objects such as a threaded cylinder \cite{straley,kimura2},
twisted rod \cite{chitin-revol,orts-cellulose} or semi-flexible helix \cite{pelcovits}.

\begin{figure}
\includegraphics[width=0.55\columnwidth]{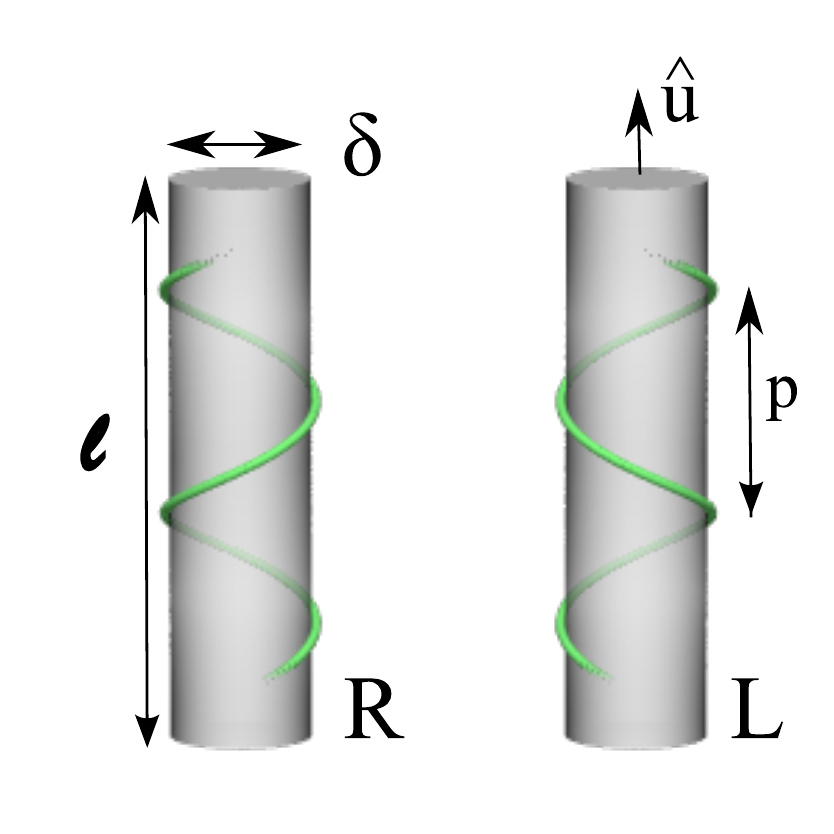}
\caption{ Cylinder of dimensions (${ \ell , \delta }$) enwrapped with a helical segment potential with internal
  pitch length $p$. The helix sense can be right-handed (R) or left-handed (L). }
\label{fig1}
\end{figure}

Despite recent progress in the simulation domain \cite{tombolatoferrarini, frezza2013} a common theoretical framework capable of rationalising the pitch trends of cholesteric materials starting from the microscopic properties of the constituents remains elusive. 
In this paper we endeavour to make a first step in this direction by considering a semi-analytical density functional treatment of cholesteric assemblies starting from a generic helical segment model.
To establish a microscopic understanding of the subtle connection between micro- and macrochirality we start by deriving the effective chiral potential between two slender helical objects as a generic model for chiral nanoparticles with arbitrary internal helicity.  Next,  the implications of such  chiral
potentials on the structure
and symmetry of a cholesteric phase  will be addressed using statistical mechanical theory. 
Our chiral potential has a simple pseudo-scalar form similar to the ones derived from more explicit electrostatic models in which chiral interactions are mediated through helically arranged local dipoles  \cite{goossens}. Owing to its tractable form the pseudoscalar chiral potential is routinely invoked in particle simulations of cholesteric mesophases \cite{berardi_zannoni1998}.  
It can also be combined with a Maier-Saupe
mean-field theory \cite{meervertogenJCP,lin-liu77}, or with a bare
hard-core model and treated with a virial theory
\cite{onsager,vargachiral1,wensinkjackson} to  study the structural properties of the cholesteric phase. In this work we shall use
an extended Onsager theory, due to Straley \cite{straley},  as a microscopic framework to assess the pitch sensitivity with respect to the helical properties of the constituents as well as the thermodynamic state of the system.  

The magnitude and symmetry of the cholesteric pitch turn out to be sensitive to not only the
microscopic pitch but also the degree of alignment along the helical director field. The latter, in turn, can be steered by the density (lyotropics), temperature (thermotropics) or by some directional external field. To illustrate the concept, we show that a helix with fixed internal pitch may self-assemble into both right- and left-handed chiral phases whose handedness may spontaneously switch depending on the thermodynamic state of the system. Such pitch inversions have been found in various experimental observations \cite{watanabe,yamagishi} but a sound statistical mechanical underpinning of these phenomena is lacking 
mainly because the construction of generic, predictive models is strongly impeded by the complicated physico-chemical nature of many thermotropic liquid crystals. 

\section{Model} 

Let us consider a pair of strongly elongated  helices each described by a linear array of rigidly linked soft segments with a radially symmetric  interaction potential $u_{s}(r)$ wrapped around a cylindrical backbone [see \fig{fig1}]. In the continuum limit the potential $U_{h}$ between two helices with length $\ell$ depending on the centre-of-mass distance $\bfr_{12}$ and solid orientation angles $\Omega_{i}$ formally reads:

\beq
\label{segpot}
U_{h}  =  \int _{-\ell/2}^{\ell/2} dt_{1} \int_{-\ell/2}^{\ell/2} d t_{2}  u_{s}( |  \bfr_{12} + \bs_{1}- \bs_{2} | ),
\eeq
where $\bs_{i}(t_{i}, \oma)$ denotes the local segment position of rod $i$  parameterized by $t_{i}$.
A helix of diameter $\delta$ can be defined by invoking a molecular orthornormal basis $\{ \bhu_{i} , \bv , \bw_{i} \}$ $(i=1,2)$ in terms of the longitudinal orientation vector $\bhu$ and auxiliary unit vectors  $\bv = \bhua \times \bhub  / |\bhua \times \bhub |$ and $\bw_{i} =
\bhu_{i} \times \bv $.   The contour vector of helix $i=1,2$ then takes on the form

\beq
 \bs_{i} = \bfr_{i} + \frac{t_{i}}{2}   \bhu_{i} +  \frac{\delta}{2}  \left \{  \cos ( q t_{i}  + \az_{i} ) \bv + \sin ( q t_{i} + \az_{i} ) \bw_{i} \right \},
 \eeq
with $q = 2\pi / p$ the internal helical pitch such that $q>0$ corresponds to a right-handed (R) helix and $q < 0$ to a left-handed (L) one. 
Since a helical object is {\em  not} invariant with respect to rotations about its longitudinal axis $\bhu_{i}$ the pair potential must  explicitly depend on a set of (internal) azimuthal angles $0 \le \psi_{i} \le 2 \pi$.  To derive a simple expression for the chiral
potential  associated with the rather intractable form \eq{segpot} we follow the procedure outlined in an earlier paper \cite{wensink_jpcm2011}. First, we focus on 
strongly elongated  helices  and expand $U_{h}$ for small width-to-length ratio $\delta/\ell \ll 1$. The leading order term is of ${\mathcal O}((\delta/\ell)^{2})$ and embodies all chiral contributions for slender helices. The next step is to mitigate the multi-angular dependency of $U_{h}$  by constructing an {\em angle-averaged} chiral potential $\bar{U}_{c}$ obtained by preaveraging over the internal azimuthal angles. To this end we impose  the Helmholtz free energy of the
angle-averaged potential to be equal to that of the full
angle-dependent potential \cite{israelachvili}. Setting the thermal energy $k_{B}T$ to unity we may write the potential of mean force  in the following way:

\beq
 { \bar U_{c}}   =    - \ln  \left \langle   \exp [ -   U_{h} ]  \right \rangle _{\psi}  =   \left \langle  U_{h}   \right \rangle _{\psi}   -\frac{1}{2}  \left \langle  U_{h} ^{2}  \right \rangle _{\psi}  +  \cdots,
 \eeq
where the brackets denote a double integral over the internal angles $\langle . \rangle _{\psi}  = (2 \pi)^{-2} \int_{0}^{2 \pi} d \aza d \azb $.  
The last term can be identified with the strength of the {\em azimuthal fluctuations} and is obtained by expanding the free energy up to quadratic order in 
 $U_{h} $. It can be readily shown that the simple average yields zero ($ \langle   U_{h} \rangle _{\psi} =0 $) so that only the quadratic fluctuation term survives.  This is consistent with the notion that the azimuthal helix-helix correlations play a key role in stabilising cholesteric order, as discussed in \cite{harris-rmp}. 
Physical justification of the expansion above relies on the observation that in most experimental systems the cholesteric twist deformation is weak  ($\gg \ell $). As a result, the chiral contribution to $U_{h}$ which is the only part responsible for the formation of a helical director field is generally much smaller than the thermal energy. The integrations over the azimuthal angles are
trivial and all contributions invariant under a parity transformation $ \bfr_{12}
\rightarrow - \bfr_{12}$ are non-chiral and may be discarded.  Combining all relevant contributions leads us to the following compact expression for the chiral potential
between two strongly elongated helices with $\delta /\ell \gg 1$:

\beq 
\bar{U}_{c} (  \bfr_{12} ; \bhua , \bhub )  \simeq 
\frac{1}{4} \left ( \frac{\delta}{\ell} \right ) ^{2}  {\cal F} (r_{12} , q ) (\bhua \times \bhub \cdot \ell^{-1} \bfr_{12} ).
 \label{uchi}
\eeq
The term between brackets is a chiral pseudo-scalar which changes sign under a parity transformation and is routinely imposed  to describe chiral interactions \cite{goossens}. We show that this form naturally emerges as the leading-order chiral potential for slender helical objects. Most importantly, however, our prefactor provides direct access to the microscopic helical pitch via: 

\beq
{\cal F}( r_{12} , q)  =     \langle u_{s}^{\prime}( \tilde{r}_{12})\cos  q \ell \ta \rangle_{t}  \langle  u_{s}^{\prime}( \tilde{r}_{12}) \tb \sin  q \ell  \ta \rangle_{t},
 \label{fdr}
\eeq
in terms of the double contour average $\langle \cdot \rangle_{t}  = \int_{-1}^{1} dt_{1} dt_{2} $, intersegment force $u_{s}^{\prime}(x)
= -\partial u_{s}(x) / \partial x $ and linear segment  distance 
 $\tilde{r}_{12}^2  =  \ell^{-2} r_{12} ^{2} + \frac{1}{4} (\ta \bhua - \tb \bhub )^{2} $.  In order to appeal to both lyotropic and thermotropic assemblies of helical building blocks we may consider two different segment potentials. First, a (repulsive) Yukawa segment potential $u_{s}(r) = u_{0}\exp(-\kappa r)/r$, with $\kappa$ an inverse electrostatic screening length,  provides a relevant description of charge-stabilised colloidal helices whose self-assembly properties are governed mainly by particle concentration. To make a connection to  thermotropic systems, we consider a van der Waals (vdW) form $u_{s}(r) = -u_{0} r^{-6}$ in which case the system temperature rather than concentration constitutes the chief thermodynamic control parameter owing to the long-ranged attractive interparticle forces.  The amplitudes $u_{0}>0$ pertain to various electrophysical properties (surface charge, dielectric constant etcetera) of the individual helices which we do not need to not specify here.

\begin{figure}
\includegraphics[width=0.8\columnwidth]{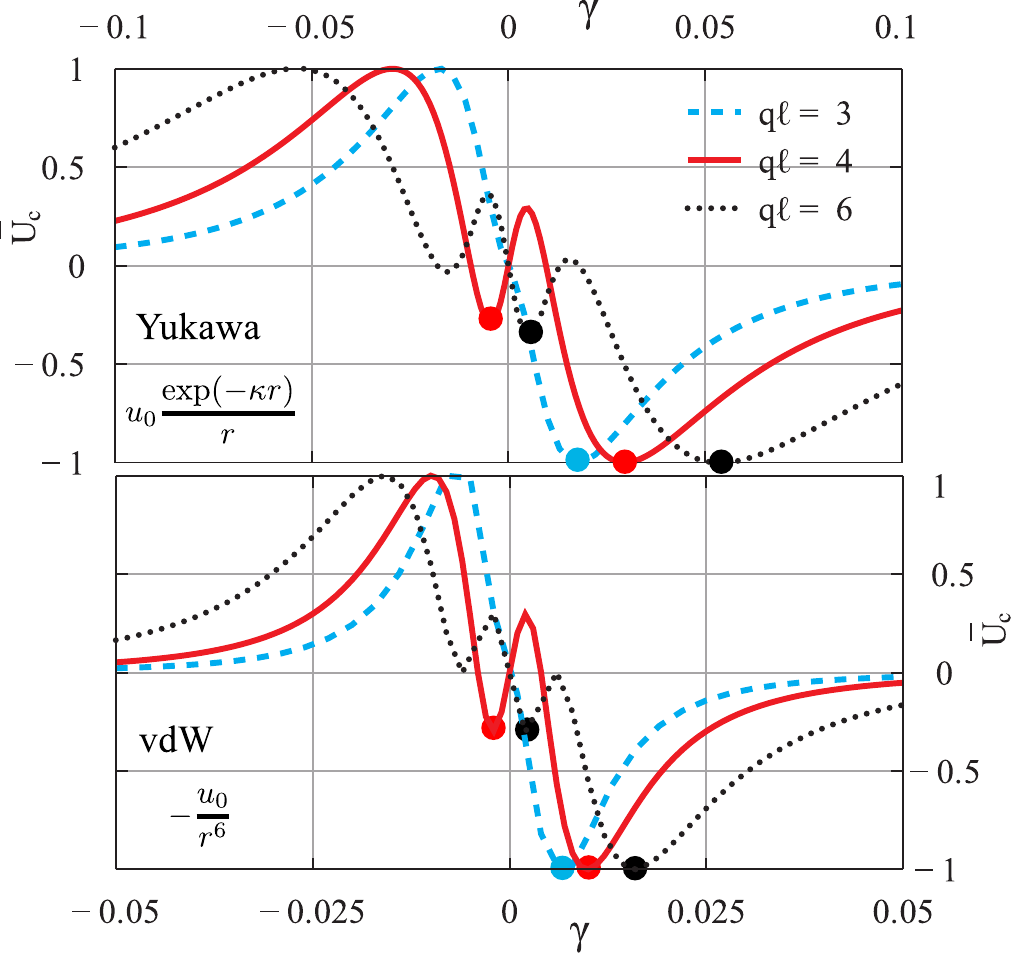}
\caption{Angular variation of the near-field chiral potential [\eq{uchi}] at fixed rod distance $0.1\ell $ depends sensitively on the molecular pitch $q$. The curve for $q \ell = 4$ reveals a double minimum at opposing twist angle $\gamma$, irrespective of the nature of the interactions as demonstrated for two different segment potentials $u_{s}$. The extrema have been scaled to unity to facilitate comparison.}
\label{fig2}
\end{figure}

Irrespective of the nature of the segment potential the chiral potential exhibits an intricate angular dependence (\fig{fig2}). Results are shown for a particular interhelix distance of $0.1 \ell$ but the overall 
 features do not change qualitatively for different values, provided the distance remains larger than the core diameter $\delta$. 
 In particular, the amplitude and direction of the effective torque each helix experiences depends sensitively on its local orientational freedom. Since the latter is tuned primarily by density or temperature we expect a highly non-trivial response of the cholesteric symmetry upon variation of these quantities.  Two observations in \fig{fig2} hint at a subtle relationship between the helical properties of the individual particles and those of the macroscopic phase. First, for $q=4$ the sign of the effective torque $ \tau \sim -  (\partial \bar{U}_{c}/\partial \gamma )_{\gamma = 0}$ at small mutual angle $\gamma(\bhua, \bhub)$  is opposite to that of the other helices shown. This implies that helices may stabilise a cholesteric helix sense with {\em opposite} symmetry  in the asymptotic limit of strong alignment ({\em viz.} very large concentrations) \cite{wensinkjackson}.  A second, more implicit, observation is that for certain values of $q$, the local and global minima correspond to {\em opposite} torque directions.  The consequence is that the symmetry of the effective microscopic torque experienced by each helix due to correlations with its neighbours  depends crucially on the degree of local alignment along the helical director.  

\section{Onsager-Straley theory} 

To scrutinise the effect of these subtleties on the macroscale we invoke a simple Onsager-type theory appropriately generalised for weakly helical director fields with pitch $k \ll \ell^{-1}$  \cite{straleychiral,allenevans}. The Helmholtz free energy density $F$ per unit volume $V$ depends on the one-particle orientational distribution $f(\bhu)$  reads up to quadratic order in  $k$:

\beq
\frac{F}{V}  = 
\rho  \int d \bhu f (\bhu ) (\ln [\rho {\mathcal V} f( \bhu ) ]-1)  + \sum_{n=0}^{2}  K_{n} (-k)^{n} /n !,  \label{freec}
\eeq
with $\rho$ the particle number density and ${\mathcal V}$ the immaterial thermal
volume of a helix. \heq{freec} reflects a  balance between the ideal mixing and (local) orientational 
entropy and the excess free energy accounting for helix-helix interactions on the
second-virial level in terms of the following angular averages \cite{allenevans} 

\beq
K_{n}[f]  =   \frac{\rho^2}{2} \int d \bhua  \int d \bhub  [ \partial_{n} f (\bhua)
f (\bhub) ]  M_{n} (\bhua , \bhub ),  \label{twistk}
\eeq
in terms of the derivatives $\partial_{0} =1$, and

\bea
\partial_{1} &=& u_{2\perp} \partial_{\bhub}, \nonumber \\ 
\partial_{2} &=& u_{1\perp} \partial_{\bhua} u_{2\perp} \partial_{\bhub},
\eea
acting on $f$ with $( \parallel, \perp )$ denoting the vector component along and transverse to the cholesteric pitch  direction. 
The reference term, $K_{0}$ is associated with an untwisted
nematic system,  whereas $K_{1}$ embodies an effective torque-field emerging from the chiral potential.  $K_{2}$ represent a twist elastic energy counteracting the helical deformation of the director field.  The  kernels  \eq{twistk} are entirely microscopic and are given by higher-order spatial averages of the Mayer function of the helical pair potential

\beq
M_{n} =  - \int d \bfr_{12}   r_{12 \parallel} ^{n} (e^{-U_{h}} -1).  
\eeq
If we assume helix envelope to consist of a cylindrical hard inner core of diameter $\delta$, then  

\beq
M_{0} = 2\ell^{2} \bar{\delta} | \sin \gamma |,
\eeq
identical to the  excluded volume $v_{\rm{ex}}$ of the cylinder-shaped helical envelope.  
The soft potential can be  subsumed into an effective, angle-dependent diameter $\bar{\delta} = \varepsilon(\gamma) \delta $. The orientation-dependent prefactor reads

\beq
 \varepsilon(\gamma) = 1 +  \int_{1}^{\infty} d x (1- \exp[ - u_{s}(x) \cos^{2}(\gamma) ] ). 
 \eeq
which reduces to unity for strictly hard rods ($u_{s} = 0$). The cosine term reflects the intrinsic tendency of attractive helix pairs to align and repulsive ones to adopt a perpendicular configuration \cite{stroobantslading}. Similar arguments can be applied to the twist elastic constant in which case the kernel is represented by some higher-dimensional excluded volume 

\beq
M_{2} = \frac{1}{6} \ell^{4} \bar{\delta} | \sin \gamma | (u_{1\parallel}^{2 } +  u _{2\parallel}^{2}).
\eeq
The symmetry of $M_{1}$ dictates that the torque-field constant $K_{1}$ depend only on the pseudo-scalar contribution ro the helix potential [\eq{uchi}].  Recalling that $ \bar{U}_{c} \ll 1$  and adopting a simple van der Waals ansatz one arives at a tractable form

\beq
 M_{1} \simeq \int _{\notin v_{\text{ex}}} d \bfr_{12} r_{12 \parallel}   {\bar U}_{c}(\bfr_{12}, \bhua, \bhub),
 \eeq 
where the spatial integral runs over the space complementary to the excluded volume $v_{\rm{ex}}$ of the helix envelope.

\section{Pitch inversion} 
Most helically organised assemblies known in experiment  possess a pitch length  much larger than the molecular size.  It is therefore reasonable to suppose that the  local  nematic order is only marginally affected by the twisted director field. In this situation the local orientational distribution of the main helix axis $f(\bhu)$ can be established from a formal minimisation of the nematic free energy [\eq{freec}, setting $k=0$] so that
 
 \beq
 f(\bhu) = {\mathcal N} \exp ( - c \int d \bhu^{\prime} \varepsilon (\gamma) | \sin \gamma  | f(\bhu^{\prime})), 
 \eeq
where the constant ${\mathcal N}$ ensures normalisation and $c = \rho \ell^{2} \delta$ defines a dimensionless concentration measure.   From $f^{\ast}$ one can extract the nematic order parameter along the local director $\bn$ via $ S = \int d \bhu  f(\bhu) {\mathcal P}_{2}(\bhu \cdot \bn)$ (with ${\mathcal P}_{2} (x) = \frac{3}{2}x^{2} -\frac{1}{2}$ a Legendre polynomial). The ratio of the microscopic constants $K_{i}$ define the equilibrium cholesteric pitch 

\beq
k = K_{1}[f] / K_{2}[f]. 
\eeq
This result naturally follows from the extremum condition $\partial F /\partial k =0$ and reflects a balance between the torque-field and twist elastic contributions to the free energy.  In keeping with the internal pitch we identify $k^{\ast}>0$ with a right-handed (R) helical director field and $k^{\ast}<0$ with a left-handed  (L) one. With this, we have established the desired connection between thermodynamic variables (concentration or temperature) and cholesteric pitch $k^{\ast}$ for helical particles with arbitrary internal pitch $q$.

\begin{figure}
\begin{center}
\includegraphics[width= 0.8 \columnwidth]{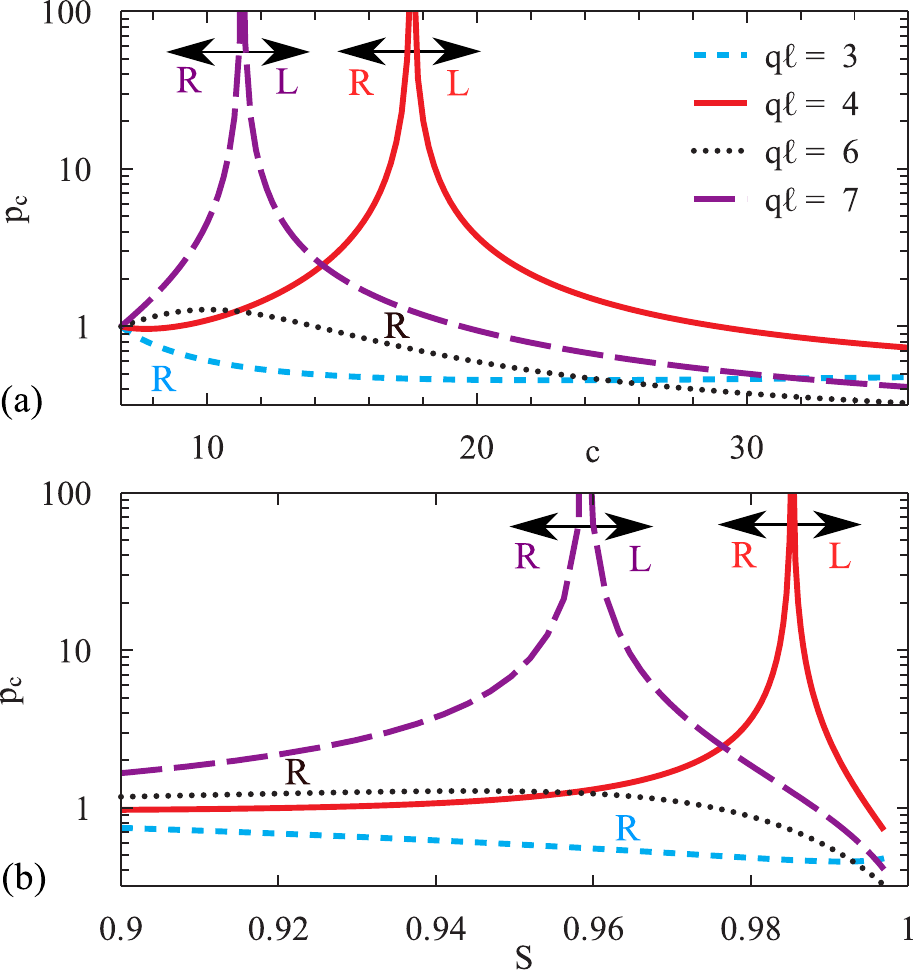}
\caption{ \label{fig3} (a) Cholesteric pitch length $p_{c}$ versus concentration for a system of helical Yukawa rods with $\kappa \ell =
  20$ and $\delta/\ell = 1/50$ for different values of the internal pitch  $q \ell$.  
   (b) Same result plotted against the local nematic order parameter $S$.  A pitch sense inversion (right-handed $\leftrightarrow$ left-handed) occurs for $q \ell = 4$ and $q \ell = 7$.  }
\end{center}
\end{figure}

To illustrate the pitch sensitivity of cholesteric assemblies we now focus exclusively on lyotropic cholesterics composed of Yukawa helices for which the concentration $c$ constitutes the main thermodynamic parameter. 
The results in  \fig{fig3} show the variation of the cholesteric pitch length with $c$ for  different values of the internal pitch $q$.  The cholesteric pitch has been normalised to its value corresponding to  the cholesteric phase at coexistence with the isotropic phase ($c=6.28$ setting $u_{0}=1$) to avoid having to make an explicit reference to the physico-chemical helix details that go into $u_{0}$ \cite{wensink_jpcm2011}. 

The helices corresponding to \fig{fig3} all possess a right-handed
 symmetry and one would naively expect the  cholesteric phase to
 adopt the same symmetry. Fig. 3 shows that this is indeed the case for
  $q\ell = 3$ and $q\ell=6$ where the cholesteric sense remains right-handed (R) throughout the probed concentration
 range, but not for $q\ell=4$ and $q \ell =7$.  In the latter cases a more complicated scenario if found in which a R-cholesteric phase
transforms  into a L-phase upon increasing $c$. The critical value at which the
sense inversion occurs is found to be $c \approx 17.6 $ for the weakly coiled
 ($q \ell = 4$) and  $c \approx 11.4$ for the strongly coiled ones
($q \ell =7$). The transition from R to L is continuous and must be associated with a diverging pitch length $p_{c}
\propto |c-c^{\ast} |^{-1}$ at the inversion point $c^{\ast}$ where the system becomes nematic. 
For $c< c^{\ast}$ the pitch strongly decreases upon lowering $c$ and a
distinct unwinding of the helical director field occurs close to the transition
towards the isotropic phase. Symmetry prescribes the same sequence
of pitch changes to occur for {\em left-handed} helices with the sense changing
R $ \rightarrow$ L upon dilution.
Independent of $q$ the cholesteric becomes more strongly coiled upon increasing  concentration and the pitch length attains a simple proportionality  $p_{c} \propto 1/c$  in the asymptotic concentration limit \cite{odijkchiral}.
 
\begin{figure}
\begin{center}
\includegraphics[width= 0.8 \columnwidth]{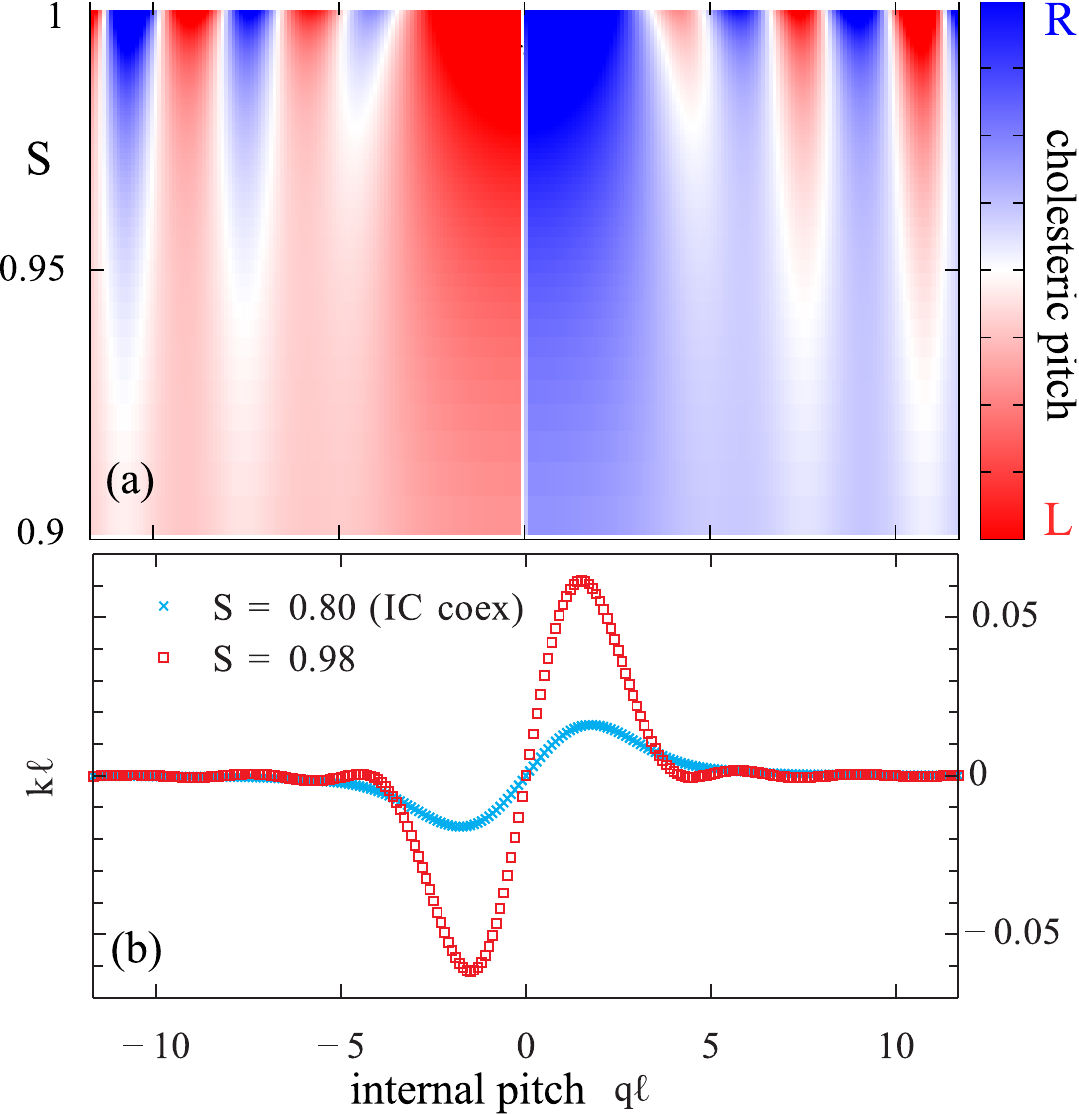}
\caption{ \label{fig4} (a) Relation between the cholesteric pitch $k$ and internal pitch $q$ as a function of the degree of local nematic order $S$. The sign and strength of the cholesteric pitch (in reduced units) is indicated by colour coding. The white zones refer to nematic regions ($q =0$) connecting the two helical senses. (b)  Absolute value of the cholesteric pitch $k \ell$ for two values of the local nematic order parameter.  }
\end{center}
\end{figure}

 \heq{fig4} presents a schematic overview of the interrelation between microscopic and cholesteric chirality. We can infer that pitch inversions upon change of local nematic alignment only occur  in certain $q$ interval while absent in others.   The pitch amplitude  (\fig{fig4}b) depends sensitively on  $q$ with the pronounced extremum around $| q | \ell \sim 3$ revealing an optimal `twisting strength'  for moderately coiled nanohelices \cite{wensink_jpcm2011}.
As alluded to in \fig{fig2}, the sense inversion is 
imbedded in the intricate dependence of the chiral potential on 
the microscopic twist  angle $\gamma$.  A prerequisite for the pitch inversion is the presence of an {\em antagonistic} effect in the azimuthally averaged interhelix potential 
represented by minima located at opposite sign of the angle $\gamma$ between the main helix axes.
The ratio at which these minima are sampled depends crucially on the degree of nematic alignment around the local director and a  change of nematic order (by varying particle concentration or temperature)
allows the helix pairs to preferentially adopt either a positive or negative twist which then proliferates towards the formation of a left- or right-handed director field.

In view of the similarity between the scenarios depicted in \fig{fig2} one can envisage an analogous pitch inversion for  attractive van der Waals segment potentials. This situation would correspond to  thermotropic helical assemblies where a change of temperature $k_{B}T/u_{0}$ (at fixed pressure) provides the main driving force for liquid crystal order. 
The present theory could therefore also be used to model thermotropic systems of coiled molecules in which a similar complex interplay between micro- and macrochirality  can be expected by variation of temperature.

\section{Conclusion} We propose a course-grained helical  segment model 
to study  chiral self-organisation in lyotropic or thermotropic assemblies of helical mesogens. 
From the general pair potential we extract  an algebraic chiral potential similar to the pseudoscalar form
\cite{goossens} widely used to describe
long-ranged chiral dispersion forces. Whereas the pseudoscalar model potential usually requires an unknown adjustable prefactor, our agebraic form provides explicit reference
to the molecular helicity.
By combining the potential
with a simple Onsager-Straley theory we study the cholesteric pitch as a function of the magnitude and sense of the
pitch as well the thermodynamic state.  The cholesteric handedness is not a priori dictated by the symmetry of the individual helices but 
depends sensitively on the precise value of the internal pitch and the thermodynamic state of the system. We map out the precise conditions under which right-handed 
helices generate left-handed chiral phases and vice versa.
 The antagonistic effect of helical interactions is consistent with experimental observations in {\em M13} virus systems \cite{tombolato-grelet} and
various types of DNA \cite{livolantDNAoverview,tombolatoferrarini}
where left-handed cholesteric phases are formed from right-handed
helical  polyelectrolyte conformations.
Small variations in the shape of the helical coil,
induced by e.g. a change of temperature, may lead to a sense inversion
of the helical director. Such  inversions have been found in thermotropic (solvent free) polypeptides \cite{watanabe}, cellulose derivatives \cite{yamagishi}, and in mixtures of right-handed cholesterol chloride and left-handed cholesterol myristate \cite{sackmann}.

The present model could be interpreted as a benchmark for
complex biomacromolecules such as DNA and {\em fd} which are characterised
by a helical distribution of charged surface groups. Other lyotropic cholesteric systems, such as cellulose and chitin
microfibers in solution could also be conceived as charged rods with
a twisted charge distribution \cite{chitin-revol}.  A more accurate description of the
pitch sensitivity,  particularly for DNA systems, could be achieved
by taking into account the steric contributions associated with the
 helical backbone of the chains as well as the influence of chain
 flexibility. This could open up a route towards understanding the unusual
 behaviour of the pitch versus particle and salt  concentration
 as encountered in DNA \cite{livolantDNAoverview,strey-dna, zanchetta_pnas2010} using simple coarse-grained models. 

Last not least, in view of their intricate interplay between micro- and macrochirality  assemblies of helical particles could be exploited for photonic applications  as well as the design of opto-electronic switching devices with improved performance and controllability \cite{kopp_2003}.

\bibliographystyle{apsrev}
\bibliography{rik}

\end{document}